# A Conceptual Blend Analysis of Physics Quantitative Literacy Reasoning Inventory Items

Suzanne White Brahmia  
University of Washington

Alexis Olsho  
University of Washington

Andrew Boudreaux  
Western Washington University

Trevor Smith  
Rowan University

Charlotte Zimmermann  
University of Washington

*Mathematical reasoning flexibility across physics contexts is a desirable learning outcome of introductory physics, where the "math world" and "physical world" meet. Physics Quantitative Literacy (PQL) is a set of interconnected skills and habits of mind that support quantitative reasoning about the physical world. The Physics Inventory of Quantitative Literacy (PIQL), which we are currently refining and validating, assesses students' proportional reasoning, co-variational reasoning, and reasoning with signed quantities in physics contexts. In this paper, we apply a Conceptual Blending Theory analysis of two exemplar PIQL items to demonstrate how we are using this theory to help develop an instrument that represents the kind of blended reasoning that characterizes expertise in physics. A Conceptual Blending Theory analysis allows for assessment of hierarchical partially correct reasoning patterns, and thereby holds potential to map the emergence of mathematical reasoning flexibility throughout the introductory physics sequence.*

*Keywords:* physics, quantitative reasoning, mathematization, conceptual blend, assessment

Introductory physics is characterized by using simple mathematics in sophisticated ways. Physics equations tell stories. For example, to an expert, $a = -9.8 \ m/s_2 + -b(2m/s)$, the equation is recognized as describing the acceleration of an object that is in free fall and experiencing air resistance. The two forces causing the acceleration in this case are the gravitational force and the force of air resistance. The coordinate system is set by the choice of sign for the acceleration due to gravity, "downwards" in this case is chosen to be in the negative direction. At this instant, the object is moving upwards at 2 m/s since the one-dimensional velocity vector is in the positive direction. The negative signs in front of each of the two terms on the right-hand sign of the equation carry different meanings. The first negative sign is an arbitrary choice that determines the coordinate system, while the negative sign on the second term indicates that whatever the sign of the velocity is, this contribution to the (vector) acceleration will be in the direction that is opposite to the direction of motion. So, the sign of the velocity, and the sign of the contribution from gravity must agree with the choice of coordinate system, but the sign in front of the second term does not because it indicates opposition, which is independent of the coordinate system used. Black and Wittman provide evidence that many of these nuances are lost on physics majors at the junior level, even though the mathematics involved is at the precaclulus level, and they are well beyond that stage in their mathematics course taking (Black & Wittmann, 2009).

Even as students move beyond the introductory sequence to using newly learned mathematics (calculus, linear algebra, differential equations), there is mounting evidence that although they don't struggle to execute the mathematics, they don't really understand why they do what they do, and they'd like to (Caballero, Wilcox, Doughty, & Pollock, 2015).

In the work described in this paper, our intention is to understand and assess the nature of student reasoning about the quantitative models that make up introductory physics, as the air resistance equation above exemplifies. Sherin's Symbolic Forms provides a framework of the kind of reasoning most physics instructors would like to see in their students as the outcome of having taken an introductory physics course (Sherin, 2001). While Symbolic Forms emerge from observations of students engaged in successful problem solving, the students in Sherin's study are high achieving students in their last semester of introductory physics at an elite institution. Most students who take introductory physics are less sophisticated mathematically and come from less educationally privileged backgrounds. We consider Symbolic Forms as a *learning objective* of the introductory physics course.

In the next sections, we describe the Physics Inventory of Quantitative Literacy (PIQL), which is designed to help assess this learning objective, and Conceptual Blending Theory (CBT), which provides a framework for understanding the integration of mathematical and physical reasoning (Fauconnier & Turner, 2008). The purpose of this paper is to examine items from the PIQL through the lens of CBT. We argue that CBT analysis lends itself to the development of an inventory that promotes a growth mindset associated with mathematical reasoning in physics.

### Physics Inventory of Quantitative Literacy (PIQL)

Physics, as perhaps the most fundamental and the most transparently quantitative science discipline, should play a central role in helping students develop quantitative literacy (Steen, 2004). We coin *Physics Quantitative Literacy (PQL)* to refer to the rich ways that physics experts blend conceptual and procedural mathematics to formulate and apply quantitative models. Quantification is a foundation for PQL, using established mathematics to invent and apply novel quantities to describe natural phenomena (Thompson, 2010; Brahmia, 2019). These quantities then allow for the investigation of patterns and relationships, which in turn anchor the quantitative models that are the hallmark of physics as a discipline. PQL, involving sophisticated use of elementary mathematics at least as much as elementary applications of advanced mathematics, is more challenging for students than many instructors may realize (Rebello, Cui, Bennett, Zollman, & Ozimek, 2007; Brahmia & Boudreaux, 2016).

Although the mathematics involved in introductory physics quantification is typically algebra or arithmetic, a conceptual understanding of this mathematics is fundamental to the sophisticated task of reasoning in the context of strange new physics quantities. Despite its importance, little work had been done to measure progress of PQL in introductory physics students as a result of instruction. The Physics Inventory of Quantitative Literacy (PIQL) is a valid and reliable reasoning inventory that is under development by the authors with the intention of being used to track changes in student quantitative reasoning over the course of the introductory physics sequence. Carefully validated forced-response questions probe student reasoning about quantification in introductory-level physics contexts. Analysis of PIQL results allows us to track progress in students' PQL and determine features of PQL that are particularly challenging.

We've identified three facets as the basis of quantification in introductory physics: proportional reasoning, reasoning about signed quantities, and co-variational reasoning. The choice of these three facets as foundational for physics quantitative literacy was supported by work done in both physics education research and mathematics education research. Much of our thinking about the domain of co-variational reasoning originates in mathematics (Carlson, Oehrtman, and Engelke, 2010). The development of the Pre-calculus Concept Assessment served as a foundation for our thinking about covariation in physics.

Concept and reasoning inventories, by nature, use expert-like reasoning and understanding as a metric with which to assess student reasoning. In the process of developing items for the PIQL, we developed an organizational framework by which to characterize the different uses of the negative sign in contexts of introductory physics (Brahmia, Olsho, Smith & Boudreaux, 2018). This framework proved useful not only in the development of assessment items for the PIQL, but also, potentially, for instructors and researchers in characterizing student understanding of the different meanings of the negative sign in physics contexts. We continued work on a framework for the natures of proportional reasoning in introductory physics (Boudreaux, Kanim & Brahmia, 2015) and have begun work for an analogous framework for co-variational reasoning.

In its current state, the PIQL items (including the distractors) are representative of expert natures of proportional reasoning, co-variational reasoning and reasoning with signed quantities. The item distractors emerge from open-ended version analysis of student responses, which are then refined to align with validated expert natures. We conduct student think-aloud interviews to refine or reject items. We use CBT (Fauconnier & Turner, 2008) as a framing for the analysis of student responses to help inform both item refinement and implications for instruction.

Results from the Physics Inventory of Quantitative Literacy (PIQL) administered in large-enrollment calculus-base courses - when the items are scored dichotomously (either all correct or incorrect) - indicates only a very modest improvement in students' quantitative reasoning as a result of introductory-level physics instruction. By incorporating patterns that emerge from a CBT analysis, we are encouraged at the potential to recognize hierarchy in students "incorrect" answering, and thereby be able to better understand, and assess, the development of PQL over the course of the introductory physics sequence.

## Conceptual Blending Theory

Conceptual Blending theory (Fauconnier & Turner, 2008) describes a cognitive process in which a unique mental space is formed from two (or more) separate mental spaces. The blended space can be thought of as a product of the input spaces, rather than a separable sum. We view the development of expert mathematization in physics occurring not through a simple addition of new elements (physics quantities) to an existing cognitive structure (arithmetic), but rather through the creation of a new and independent cognitive space. This space, in which creative, quantitative analysis of physical phenomena can occur, involves a continuous interdependence of thinking, most of which is subconscious, about the mathematical and physical worlds.

The following are elements found in any blend (see Fig. 1a). This static diagram represents connections that activate and deactivate. It is not depicting actual stages in a temporal way.

- Input spaces: contains the concepts involved
- Generic Space: structure that the inputs share, and maps onto input spaces
- Blended Space: related to generic space but contain more structure in which the inputs are indistinguishable
- Projections: represented by lines connecting the rectangles; project in either direction

As an example of CBT analysis, we present an abridged version of their Buddhist Monk problem. The reasoning is abstract and cognitively complex; its analysis is shown in Fig. 1(b).

> A monk begins at dawn one day walking up the mountain, reaches the top sunset, stays several days then he begins at dawn to walk back and reaches the bottom at sunset. Make no assumptions about his starting or stopping or about the pace of the trips. Is there a place on the path that the monk occupies at the same hour of the day on the two separate journeys?

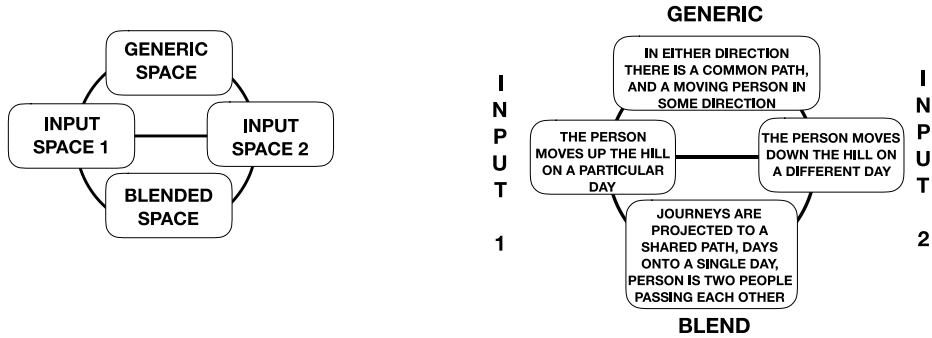

*Figure 1. (a) Elements of a typical conceptual blend, (b) CBT analysis of Buddhist Monk problem*

## CBT Analysis of PIQL Items

In what follows, we present a CBT analysis of Alex and Jessie solving two PIQL proportional reasoning items to illustrate our method. We use dotted lines in the CBT analysis to indicate unstable knowledge.

### PIQL Item 1

A miner is trading steel for lead. According to the current exchange rate, each kilogram of lead is worth 1.6 kilograms of steel. The miner trades $M$ kilograms of steel. Which of the following expressions helps figure out how many of kilograms of lead the miner will get?

a. $M \cdot 1.6$    b. $M/1.6$    c. $1.6/M$    d. $1/(1.6 \cdot M)$    e. None of these are helpful.

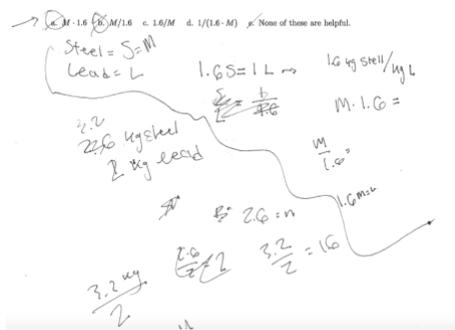
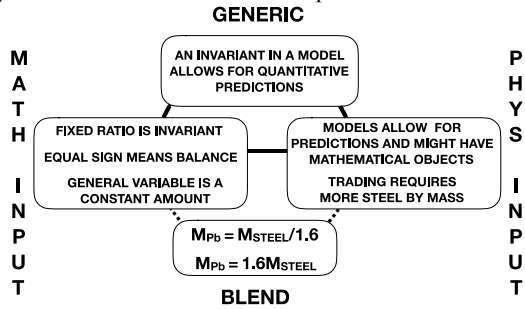

*Figure 2. (a) Alex's work showing multiple attempts to make sense of the general variable "M", (b) CBT analysis*

Although Alex's reasoning, when considered dichotomously as correct or incorrect, is in fact incorrect, there are resources used that are productive here. Alex recognized that they were looking for an invariant quantity, but their reasoning broke down when trying to reason with a general variable. In physics, we often pose questions in which the value of a quantity is represented by a general variable. Its value doesn't change during the problem, but it can take on any value. Many students, like Alex, struggle to activate mathematical reasoning that is familiar to them when they are asked to reason with quantities represented as general variables. There is weakness in the projection between the blended space and the input spaces. The generic space is stable though, and can be considered a resource.

### PIQL Item 2

*Interviewer*: Describe your thinking as you find a solution to this problem.

*Jessie*: Okay, so in this problem, we essentially have a triangle and we know two sides of the triangle and can find the third. And if the third side of the triangle, like the hypotenuse is longer, then it'll be like less steep. It's like the taller it is and then narrower the base, the steeper the actual slope of the slide. And you can just Pythagorean theorem this.

You are purchasing a slide for a playground and would like to get the steepest one you can find. For four different slides, you have the measurements of the length of the base of the slide (measured along the ground), and the height of the slide.

You decide to use this information to rank the slides from **most steep** to **least steep.** Which of the following choices is the best ranking?

| Slide | Base  | Height |
|-------|-------|--------|
| A     | 8 ft  | 12 ft  |
| B     | 5 ft  | 9 ft   |
| C     | 6 ft  | 9 ft   |
| D     | 12 ft | 8 ft   |

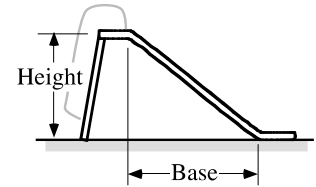

a. A > B = C > D
b. B > C > A > D
c. A = B > C > D
d. B > A = C > D
e. A = D > C > B

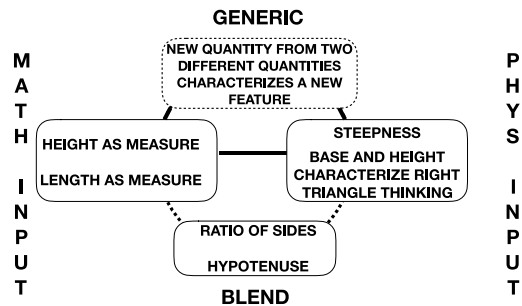

*Figure 3. CBT analysis of Jesse's reasoning*

The most common student responses to this question are shown in the blend in Fig. 3. Jessie activates the generic space shown in Fig. 3, but it is unstable; they aren't creating an independent quantity. Jessie shows weakness in the projection process between the blend and each of the two input spaces. They never reconcile that the hypotenuse is appropriate only if the height is held constant, in which case it is also unnecessary since the length alone would represent the steepness. In addition, they fall back on a common heuristic of a physics classroom, where a productive resource for Jessie when working with vector components is to "Pythagorean theorem this" to find a measurable quantity (e.g. speed from the components of a velocity vector).

### Discussion

This CBT analysis facilitates parsing out specifically where student reasoning is productive, and where problems lie. Invariant reasoning is a resource for Alex, and Jesse recognized the need to create a single new quantity from two others. These resources can seed future interventions. In both cases, we see the projection between the blend and the input spaces is unstable, and for Alex (and many students) quantity expressed as a general variable destabilizes reasoning. Neither student was completely wrong. The CBT analysis provides seeds on which reasoning can grow, providing a potential pathway to help students strengthen their PQL.

### Acknowledgments

We are grateful to the NSF for funding this work, DUE-IUSE #1832836.